# An empirical investigation of the Tribes and their Territories: are research specialisms rural and urban?

Giovanni Colavizza[1]
The Alan Turing Institute – United Kingdom
gcolavizza@turing.ac.uk

Thomas Franssen
Centre for Science and Technology Studies
Leiden University – The Netherlands
t.p.franssen@cwts.leidenuniv.nl

Thed van Leeuwen
Centre for Science and Technology Studies
Leiden University – The Netherlands
leeuwen@cwts.leidenuniv.nl

We propose an operationalization of the rural and urban analogy introduced in Becher and Trowler [2001]. According to them, a specialism is rural if it is organized into many, smaller topics of research, with higher author mobility among them, lower rate of collaboration and productivity, lower competition for resources and citation recognitions compared to an urban specialism. It is assumed that most humanities specialisms are rural while science specialisms are in general urban: we set to test this hypothesis empirically. We first propose an operationalization of the theory in most of its quantifiable aspects. We then consider specialisms from history, literature, computer science, biology, astronomy. Our results show that specialisms in the humanities present a sensibly lower citation and textual connectivity, in agreement with their organization into more, smaller topics per specialism, as suggested by the analogy. We argue that the intellectual organization of rural specialisms might indeed be qualitative different from urban ones, discouraging the straightforward application of citation-based indicators commonly applied to urban specialisms without a dedicated re-design in acknowledgement of these differences.

1) Introduction

A long tradition of sociological research aims to understand the differences in the organizational and cognitive structure of scientific fields [Merton, 1973; Whitley, 1984; Becher, 1989; Fuchs, 1993]. This sociological tradition was in its earlier years intimately connected with the emerging field of bibliometric methods and applications, originated in the 1960s with the work of Storer and Price [Storer, 1967; Price, 1970]. For example, the Price index has played an important role in the early sociology of science [Zuckerman & Merton, 1973; Cole, 1983] and is of continuous importance in scientometrics [Wouters & Leydesdorff, 1994; Larivière, Archambault & Gingras; 2008].

---

[1] Colavizza was in part supported by the Swiss National Fund under grants 205121_159961 and P1ELP2_168489.

However, the sociology of science and scientometrics have since the early 1980s drifted apart and attempts to reconcile them, or to reconcile the more theoretically inclined field of science and technology studies with scientometrics, have not had the desired effect [e.g. Leydesdorff, 1989; Luukkonen, 1997]. Recently, scholars have again argued for the need for interdisciplinary work bridging the sociology of science [Glaser & Laudel, 2016] or science and technology studies [Wyatt et al., 2017] with scientometrics.

We take up these calls and explore ways to bridge the sociology of science with scientometrics, using science mapping methods to operationalize a specific sociological theoretical framework. The field of science mapping has developed network methods to analyze the cognitive structure of different fields, as well as their relative interdependence [Börner et al, 2003; Boyack et al., 2005; Leydesdorff & Rafols, 2009; Börner, 2010; Leydesdorff, Hammarfelt & Salah, 2011. See Chen, 2017 for an overview]. Such network methods could be used to test sociologically informed hypotheses regarding the cognitive structure of scientific fields and differences between them [e.g. Fanelli & Glänzel, 2013].

One of the most influential sociological frameworks is presented in the work of Tony Becher [1989], who developed a conceptualization of academic territories and their tribes, initially on his own and later with Paul Trowler [2001]. Becher and Trowler argue that epistemic structures have both a cognitive and a social dimension and that communication practices of their tribes mirror (and thus reproduce) these structures. The framework has been critiqued extensively for its essentialist view, also by Trowler [Trowler, Saunders & Bamber 2012; Trowler, 2014], in particular for its assumption that knowledge claims are in essence hard or soft, pure and applied. However, fully accepting this criticism, their work employs inventive distinctions, such as those of convergent versus divergent fields and of rural versus urban territories. The latter is the focus point of our analysis, as it proposes to jointly explain the social and intellectual organization of different research specialisms mainly via the axes of research topics and the amount of researchers each topic gathers. Essentially, the metaphor of the rural and urban draws on the population density of rural and urban regions. It assumes, following classical sociology [Tonnies, 1957], that when a topic becomes more densely studied (and supported financially) its social and cognitive structures change. Rural areas organize in many, small topics, thus resulting in a fragmented intellectual organization, urban specialisms instead organize into few, more populated topics. This idea has some traction in scientometrics when distinguishing between communication patterns in the mathematical and natural sciences, and the arts and humanities [Hammarfelt, 2012; 2016].

Fully acknowledging that a spectrum of possibilities exists between these two conceptual opposites, this conceptualization allows us to identify a broader and related set of characteristics that rural and urban specialisms might possess. Communication patterns, which include but are not limited to publication practices, accordingly give insight into these structures, and consequently into the differences between disciplines. Based on Becher and Trowler [2001, Ch. 6] we thus develop a number of hypotheses of what we can expect to observe in rural or urban specialisms, if indeed this theory holds true.

The main distinction between rural and urban specialism is made based on the number of topics studied within a community at a given time—low for urban specialisms, high for rural specialism—and the "people-to-problem" ratio, meaning the number of researchers involved in a research topic at any one time—high for urban specialism and low for rural specialisms. We hypothesize:

> Hypothesis 1: The number of topics being researched is high for rural specialisms and low for urban specialisms. In rural specialisms more, smaller topics are expected to be found, everything else being equal, while in urban specialism fewer, larger topics are expected to be found.

> Hypothesis 2: Rural specialisms have a low people-to-problem ratio and urban specialisms a high people-to-problem ratio.

The authors subsequently suggest that this difference has implications for publication practices. They argue that rural authors have a broader scope intellectually and move more freely between topics. As there is no clear agreement about the core problems, in each publication the argument has to be embedded explicitly in the previous literature of the specialism, across topical boundaries. Therefore, these publications are on average longer, have more references and references are more evenly distributed across the specialism, as they are less focused on the specific topic. In urban specialisms, on the other hand, publications are shorter, contain less references and references are highly specialized, as there is no need to legitimize or contextualize the publication by referencing outside of the topic. We hypothesize:

> Hypothesis 3: publications in rural specialisms are longer.

> Hypothesis 4: publications in rural specialisms contain more references, both in absolute sense and relative to the publication length.

> Hypothesis 5: references in publications from rural specialisms cover a larger variety of sources (e.g. more different journals).

> Hypothesis 6: in rural specialisms there are comparatively more core publications that are shared beyond topics, such as renown monographs, making the specialism more reliant on them overall. This is less the case in urban specialisms, where core publications are mostly restricted to a topic, such as highly cited papers introducing a new method. By core we mean relatively highly cited publications. Intuitively, there are more weak ties across topics in rural specialisms than urban ones, due to the need to embed arguments within the broader specialism and not just within the specific topic.

The rural and urban distinction also has implications for productivity and collaboration practices, although these are, according to Becker and Trowler [2001], primarily effects of higher competition in urban specialisms rather than resulting from its internal cognitive and social structure. They further argue that in a more competitive specialism, productivity is higher. Moreover, because there are many people working on similar problems, there is a heavy competition to be the first to solve research problems. In such competitive environments, there is a self-reinforcing incentive to work together, therefore the average number of authors is higher. We hypothesize:

Hypothesis 7: authors in rural specialisms publish less, but across a wider range of topics. Scholars in urban specialisms publish more but within a smaller range of topics.

Hypothesis 8: the average number of authors is higher in an urban specialism than in rural specialism, and there are more collaborations in urban specialisms.

The question why a specialism is urban or rural in the first place is a crucial one that we cannot answer in the present analysis. The authors themselves suggest that the amount of competition is the main reason for a specialism to become urban, and this can be the result of an emergent new research paradigm following a Kuhnian revolution or be influenced by science policy [Becher & Trowler, 2001: 105-106]. There are a few more aspects in the urban and rural analogy which we cannot consider in this paper: the authors suggest that urban specialisms show a stiffer competition for resources (e.g. budget allocation, students, etc.), and have more rapid and heavily used (in)formal information networks.

## 2) Operationalization

The first step into the operationalization of the rural and urban conceptualization of the structure of scientific fields is to proxy its two basic units of analysis: specialisms and topics. A specialism is a self-organizing group of people (a community) focusing on related topics of research communicated internally via specialized journals, conferences and seminars [Morris and Van der Veer Martens, 2008]. A topic of research is a well-identified set of problems and related questions, recognized by the community as being of interest and part of it. For example, in the specialism of natural language processing, we consider speech recognition a topic. Topics can be individuated at different granularities. A strong assumption is that specialisms and topics must be identically defined in order for any comparison to be cast across different fields of research.

*We proxy a specialism by considering a community producing publications which are a-priori well-individuated (by publication venue).*

*We then proxy a topic as a well-connected cluster in the bibliographic coupling network of the publications published by authors active in the specialism.*

Bibliographic coupling networks can be constructed in several ways, for example considering reference overlap or textual similarity between publications, as proxies for their relatedness. *We consider a well-connected cluster to be a connected component with a minimum edge weight on every internal edge. We thus use connected components to approximate topics.* A connected component is a sub-graph where every node is connected to other nodes by at least a path. In summary: we proxy a specialism by considering a set of externally grouped publications (in our case all publications from a journal or set of journals), a topic is then a specific connected component of such publications in the resulting bibliographic coupling network, with a minimum edge weight on every component edge. In so doing, we only aim at approximating topics. To be sure, other approaches might be considered, for example individuating topics using community detection methods or topic modelling on full texts. The main issue with these methods is that it is difficult, ultimately involving judging whether topics are indeed coherent, to arrive at comparable topics across different specialisms. The proposed method preserves the benefit of simplicity of interpretation and does not require us to judge whether a topic should be identified as such but rather assumes that overlap in references identifies similarity between publications.

In what follows we focus on networks of publications and require that specialisms possess a comparable number of publications each. An alternative would have been to consider networks of authors and require specialisms to be comparable in the number of active authors. The main reason we did not pursue this direction is the added complexity in accounting for the impact of co-authorship on bibliographic coupling networks of authors.

In this study, we focus directly on hypotheses 1 and 6, and on hypothesis 2 by implication, which we consider central to the theory and little explored in the literature. We consider hypotheses 3, 4 and 8 only using metadata, since these aspects have already been considered and largely confirmed in the literature. We do not consider hypotheses 5 and 7 since they would require different design or data. In particular, hypothesis 5 would require us to first assess the variety of venues and publication typologies, likely higher in rural specialisms than in comparably-sized urban ones. Given our operative definition of specialisms and topics, we propose to operationalize the main hypotheses derived from the rural and urban analogy as follows:

*Number and size of topics (hp. 1)*: we remove edges at increasing weight thresholds. The connectivity of the network in terms of the number and size of its connected components gives us a way to measure the relative number and size of topics. According to hypothesis 1, rural specialisms will fragment into more topics given the same weight threshold than urban specialisms, as illustrated in Figure 1.

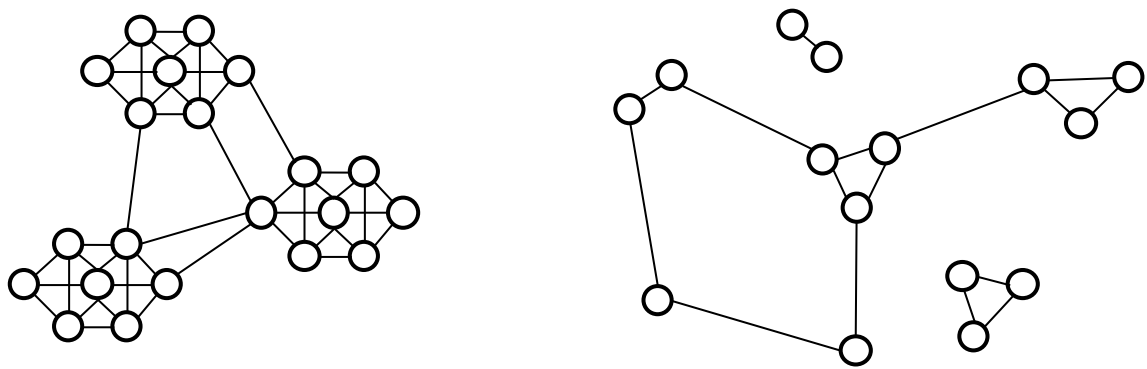

*Figure 1: Illustration of the hypothetical topic granularity in rural and urban specialisms. Urban specialisms organize tightly in fewer, larger clusters (left); rural specialisms in more, smaller clusters (right). By removing low-weight edges (represented here by longer edges), rural specialisms will disconnect into more, smaller connected components.*

*People-to-problem ratio (hp. 2)*: we operationalize the second hypothesis through the number of authors (people) active per topic (connected component) at the same edge weight threshold.

*Length of publications (hp. 3)*: we consider for each specialism the average length of publications in number of pages.

*Number of references (hp. 4)*: we consider for each specialism the average number of references, and the average number of references per page.

*Core publications (hp. 6)*: we compare the concentration of citations across the whole specialism to identify highly cited sources that we assume to be core publications. We measure the effect of core sources on the overall network connectivity by removing them in order of received citations or at random. We expect the global reliance on core sources to be greater in rural specialisms, thus the impact of removing them first to be comparatively lower on the overall network structure and size of the giant component at the specialism level. By impact here we mean the relative importance of sources into connecting the network, therefore a rural specialism will initially disconnect less rapidly by removing core sources first. Urban specialisms should be more reliant on core sources at the level of large topics, therefore they shall be less impacted by the removal of core sources as soon as the network has fragmented into topics.

*Collaboration (co-authorship, teamwork) and population (number of authors, hp. 8)*: both low for rural, high for urban specialisms. The best-known method to proxy collaborations are co-authorships. It has been already confirmed, and our study will too, that co-authorships are rarer and involving fewer authors in allegedly rural specialisms such as many in the humanities [Tsai, Corley & Bozeman, 2016].

### 3) Data

We selected ten specialisms, and corresponding datasets, within five disciplines (two specialisms each): history, computer science, astrophysics, literature and biology. Each dataset is extracted from Scopus and contains publications for every specialism over several contiguous years. These datasets are not comprehensive, something extremely difficult to achieve in general, but hopefully representative of the research published in the respective specialism. We used Scopus due to its better coverage of computer science conferences with respect to the Web of Science and its comparable coverage in general since 1996 [Harzing & Alakangas, 2016]. The selection of journals has been reviewed by at least one domain expert. International and renown venues have been preferred, as follows:

1. Specialism *A1, economic history*. Research articles from the following journals: Explorations in Economic History, Journal of Economic History, Cliometrica, Economic History Review, Business History.

2. Specialism *A2, history of science*. Research articles from the following journals: Social Studies of Science, Isis, Studies in History and Philosophy of Science, Studies in History and Philosophy of Modern Physics, Historical Studies in the Natural Sciences, Archive for History and Exact Sciences, History of Science, Annals of Science, History and Philosophy of Life Sciences, Technology and Culture.

3. Specialism *B1, computer science, neural networks and machine learning*. Conference papers from the annual conference on Neural Information Processing Systems (NIPS).

4. Specialism *B2, computer science, natural language processing*. Conference papers from the annual conference of the Association for Computational Linguistics (ACL).

5. Specialism *C1, astrophysics, solar system*. Research articles from the journal Icarus.

6. Specialism *C2, astrophysics, cosmology and astroparticle physics*. Research articles from the Journal of Cosmology and Astroparticle Physics (JCAP).

7. Specialism *D1, literature, classics*. Research articles from the following journals: Classical Quarterly, Mnemosyne, Hermes, Zeitschrift für Papyrologie und Epigraphik, Rheinisches Museum für Philologie, International Journal of the Classical Tradition, American Journal of Philology, Classical Journal, Classical Philology, Classical Receptions Journal, Quaderni Urbinati di cultura classica, Cambridge Classical Journal, Journal of Hellenic Studies.

8. Specialism *D2, English literature*. Research articles from the following journals: English Studies, Victorian Studies, Victorian Literature and Culture, Studies in English Literature, Review of English Studies, Studies in Philology, European Journal of English Studies, English Literary Renaissance, Studies in Romanticism, Journal of English Studies, International Journal of English Studies.

9. Specialism *E1, biology, neuroscience*. Research articles from the journal Neuron.

10. Specialism *E2, molecular biology*. Research articles from the journal Molecular Biology and Evolution (MBE).

The reason to select multiple journals for specialism A1, A2, D1 and D2 relates to their more diffuse publication practice (lower number of articles, higher number of book reviews per issue). As a consequence, a higher number of journals had to be selected in order to gather an overall comparable number of articles per year.

Summary statistics for the datasets under consideration are given in Table 1. The overall number of articles is comparable, yet other sensible differences emerge. In particular, articles are longer in history and literature specialisms, and they also possess fewer authors (with the partial exception of economic history), in agreement with hypotheses 3 and 8. The number of references varies greatly too, with computer science having fewer of them due to the shorter format of conference proceedings, but no clear trend is visible, contrary to what expected from hypothesis 4. In particular, there does not seem to be a clear distinction with respect to references per page, with literature possessing the lowest and biology the highest amount.

| Specialism / Statistic | A1-ec_hist | A2-hist_sci | B1-NIPS | B2-ACL | C1-icarus | C2-JCAP | D1-classics | D2-eng_lit | E1_neuron | E2_MBE |
|---|---|---|---|---|---|---|---|---|---|---|
| **Number of articles** | 1115 | 2379 | 2137 | 1904 | 2932 | 3760 | 1806 | 1159 | 2115 | 1827 |
| **Number of references** | 68'016 | 141'411 | 51'123 | 51'635 | 153'693 | 225'084 | 74'016 | 53'218 | 119'212 | 110'396 |
| **M(m) references per article** | 61.2(55) | 62.6(52) | 23.9(24) | 27.1(26) | 52.4(46) | 59.9(53) | 42.8(30.5) | 46.2(42) | 56.7(57) | 60.5(58) |

| | A1-ec_hist | A2-hist_sci | B1-NIPS | B2-ACL | C1-icarus | C2-JCAP | D1-classics | D2-eng_lit | E1_neuron | E2_MBE |
|---|---|---|---|---|---|---|---|---|---|---|
| **M(m) authors per article** | 1.8(2) | 1.2(1) | 3.1(3) | 3.1(3) | 5.2(4) | 5.4(3) | 1.1(1) | 1.1(1) | 7.5(6) | 5.5(4) |
| **M(m) pages per article** | 22.1(22) | 16.3(11) | 8.3(8) | 7.5(9) | 12.1(11) | 17.9(17) | 14.6(11) | 19(19) | 11.8(12) | 10.5(11) |
| **M(m) references per page** | 2.8(2.5) | 3.8(4.7) | 2.8(3) | 3.6(2.9) | 4.3(4.2) | 3.4(3.1) | 2.9(2.8) | 2.4(2.2) | 4.8(4.7) | 5.8(5.3) |
| **Number of articles 2016** | 147 | 350 | - | 232 | 416 | 527 | 263 | 150 | 342 | 247 |
| **Number of articles 2015** | 143 | 367 | 403 | 316 | 454 | 648 | 375 | 152 | 318 | 261 |
| **Number of articles 2014** | 137 | 357 | 411 | 286 | 431 | 658 | 229 | 160 | 343 | 277 |
| **Number of articles 2013** | 175 | 384 | 360 | 328 | 388 | 570 | 211 | 233 | 309 | 234 |
| **Number of articles 2012** | 178 | 352 | 370 | 188 | 402 | 538 | 308 | 207 | 286 | 261 |
| **Number of articles 2011** | 183 | 293 | 301 | 292 | 413 | 420 | 246 | 172 | 266 | 293 |
| **Number of articles 2010** | 152 | 276 | 292 | 262 | 428 | 399 | 174 | 85 | 251 | 254 |

*Table 1: Summary statistics for the datasets under consideration. Legend: M mean, m median. The 2016 NIPS articles were not available at the time the data was downloaded, nevertheless the year 2016 is not considered in the analysis and is only provided for reference. When references are mentioned (rows 3 and 6), all source and non-source references are considered.*

Despite the fact that the size of the selected specialisms can be considered to be comparable in terms of the number of publications, it is not so with respect to the number of authors. Table 2 reports the number of author mentions and the number of unique authors for every specialism. The number of unique authors was calculated by merging authors with the exact same surname and forename initials, assuming homonymity to be a low probability event within each specialism. Should this assumption not be valid, the number of unique authors would be higher than reported. Clearly, the number of authors publishing in each specialism varies greatly, with literature and history numerically at the bottom end, biology at the higher end, confirming hypothesis 8 in this respect.

| Specialism / Statistic | A1-ec_hist | A2-hist_sci | B1-NIPS | B2-ACL | C1-icarus | C2-JCAP | D1-classics | D2-eng_lit | E1_neuron | E2_MBE |
|---|---|---|---|---|---|---|---|---|---|---|
| Number of author mentions | 1442 | 2039 | 5685 | 4453 | 11'335 | 14'541 | 1437 | 999 | 11'321 | 7254 |
| Number of unique authors | 1107 | 1662 | 3134 | 2290 | 4808 | 5487 | 1180 | 936 | 9058 | 5552 |

*Table 2: Number of author mentions and unique authors per specialism.*

### 3.1) Data acquisition

From the Scopus interface, all relevant research articles or conference papers were downloaded, including their references. In order to include source and non-source items in our analysis merging references to the same object was need. To do so we proceeded as follows. Firstly, the authors were separated from the rest of the reference. Secondly, references without author were discarded. For two references to be merged into the same object cluster, three things need to happen: 1) the surnames of the first authors need to match; 2) the two lists of authors need to have a Jaro-Winkler score of 0.9 or above; 3) the rest of the reference text needs to have a Jaro-Winkler score above a threshold determined for each specialism/dataset. This last threshold is established empirically by finding the score yielding an accuracy of less than 0.5 in the 100 pairs of references to be merged with a score just below that threshold. Similarly, the 100 pairs immediately above the threshold must yield an accuracy above 0.5.[2] The intuition is that the accuracy of matches above the threshold rapidly improves, as it rapidly deteriorates below the threshold, therefore yielding acceptable results. The thresholds and accuracy scores for every dataset are reported in Table 3. The Jaro-Winkler measure is specifically designed to match short texts that might be misspelled at their end instead of at their beginning, such as person names. As a consequence, it is appropriate to find pairs of references similar with respect to their authors and the initial part of their title, as the rest of the reference text is more likely to contain errors or variations.

| Specialism / Statistic | A1-ec_hist | A2-hist_sci | B1-NIPS | B2-ACL | C1-icarus | C2-JCAP | D1-classics | D2-eng_lit | E1-neuron | E2-MBE |
|---|---|---|---|---|---|---|---|---|---|---|
| **Threshold** | 0.849 | 0.847 | 0.82 | 0.818 | 0.803 | 0.859 | 0.794 | 0.822 | 0.854 | 0.81 |
| **Accuracy 100 above** | 0.71 | 0.51 | 0.69 | 0.79 | 0.54 | 0.99 | 0.59 | 0.56 | 0.97 | 0.74 |
| **Accuracy 100 below** | 0.18 | 0.29 | 0.03 | 0.13 | 0.46 | 0.02 | 0.43 | 0.31 | 0.11 | 0.21 |

*Table 3: Merging threshold and evaluation of its accuracy. Some datasets have borderline results, such as*

---

[2] Accuracy is the proportion of correct matches over the total considered. An accuracy of more than 0.5 above the threshold guarantees better than chance performance. A higher threshold might be considered to rise the precision of results at a cost in terms of recall.

*Icarus and the history of science, nevertheless results rapidly improve above the threshold, as false negatives disappear below it.*

## 4) Methods

In this section we introduce two methods to construct bibliographic coupling networks for specialisms, relying on reference overlap and textual similarity. We then discuss an approach to track the number of topics of a specialism relying on the connectivity properties of these networks, and finally an approach to assess the reliance of the network on core, highly cited sources.

### 4.1) Bibliographic coupling networks

Take $B = (V, E, W)$, the weighted bibliographic coupling network [Kessler 1963] made of the publications of a specialism, in a given year or over few contiguous years. $W$ is the weighted, symmetric adjacency matrix. The edges and their weights can be established in a variety of ways. We consider two of them here: reference overlap (traditional bibliographic coupling) and textual similarity.

For reference overlap, we consider the cosine similarity over the references that two publications $i$ and $j$ have in common, and use this value to weight the edge connecting them in $B$:

$$W_{i,j} = \frac{R_{i,j}}{\sqrt{R_i}\sqrt{R_j}} \qquad (1)$$

Where $R_{i,j}$ is the number of references in common between $i$ and $j$, $R_i$ the number of references of $i$. We stress that we consider unique references, not their frequency (number of in-text references or mentions). The cosine similarity is particularly appropriate as it allows to evenly compare the weight of edges among publications with different reference list lengths, as is the case over our datasets (cf. Table 1, mean and median references per article).

We base the textual similarity among two papers on the BM25 measure, widely adopted to rank documents for the purpose of information retrieval and document clustering [Spark Jones et al. 2000a, b]. This measure has already been applied to assess the textual similarity of scientific publications [e.g. Boyack et al. 2011, Colavizza et al. 2018]. Each publication text—the concatenation of title and abstract—is reduced to lowercase and split into tokens, further eliminating punctuation and then tokens of just one alphanumeric character. Given a publication $i$ and another publication $j$, the BM25 similarity is calculated as:

$$s(i,j) = \sum_{z=1}^{n} IDF_z \frac{n_z(k_1 + 1)}{n_z + k_1\left(1 - b + b\frac{|D|}{|\overline{D}|}\right)}$$

where $n$ denotes the number of unique tokens in $i$, $n_z$ equals the frequency of token $z$ in publication $j$, and $n_z = 0$ for tokens that are in $i$ but not in $j$. $k_1$ and $b$ have been set to the commonly used values of 2 and 0.75 respectively. $|D|$ denotes the length of document $j$, in number of tokens. $|\overline{D}|$ denotes the average length of all documents in the dataset. The $IDF$ value for every unique token $z$ in the dataset is calculated as:

$$IDF_z = \log\left(\frac{N - p_z + 0.5}{p_z + 0.5}\right)$$

where $N$ denotes the total number of publications in the dataset and $p_z$ denotes the number of publications containing token $z$. $IDF$ scores strictly below zero are discarded to filter out very commonly occurring tokens. BM25 is not a symmetric measure. We obtain a symmetric measure for the similarity of documents $i$ and $j$, the value is the weight of the edge connecting them in $B$, as follows:

$$W_{i,j} = \frac{s(i,j) + s(j,i)}{2} \quad (2)$$

The BM25 textual similarity is calculated for every publication pair, independently for every dataset and within a given year interval. We refrain from further normalizing similarity scores, to allow for comparisons across datasets.

*4.2) Connectivity and giant component*

A connected component of $B$ is a sub-graph whose nodes are all connected, i.e. there exists a path between every pair of nodes in the component. An isolated node is an individual connected component. The giant component is the largest connected component measured by the number of nodes it contains [Newman 2010, 142-3].

In order to explore the connectivity property of different specialisms, using our two bibliographic coupling networks introduced above, we measure the proportion of connected components over the total possible, and the proportion of nodes in the giant component, at steps in which we remove all edges below a certain weight threshold $t$. This procedure can be considered as an analysis of a form of $t$-edge-connectivity, where a component is considered as connected only if it is a connected component by considering edges of weight at least equal to $t$. This method allows to assess the strength of edge weights in the network, and the behaviour of the connected components as the network becomes increasingly disconnected. Given our operationalization of topics, the method allows to compare the number and size of topics at different granularities, across specialisms.

In practice, given an edge weight threshold $t$, we are interested in two measures, calculated at increasing $t$ over networks $B$:

$$c(t) = \frac{C^t}{N} \quad (3)$$

$$g(t) = \frac{G^t}{N} \quad (4)$$

Where $N$ denotes the number of publications, or nodes in the specialism network, which is also equal to the maximum number of connected components in the disconnected network; $C^t$ denotes the number of connected components after removal of edges with weight strictly below $t$; $G^t$ denotes the number of nodes in the giant component after removal of edges with weight strictly below $t$.

*4.3) Core literature*

A complementary view on the granularity of topics in different specialisms can be given by considering the connectivity properties of the reference overlap bibliographic coupling network when removing highly cited sources (cf. hypothesis 6). The network will fragment after the removal of a proportion of highly cited sources, but it will do so at different speeds and times. Crucially, the more the specialism globally relies on shared sources (i.e. cited across topics), the less rapidly the network will initially fragment during such process; the more the specialism topically relies on core sources (i.e. cited within topics), the less rapidly the network will fragment once topics have been reached during such process.

We compare two processes considering the directed citation network of a specialism: one where we remove increasing fractions of cited sources in reverse order by the number of citations they received (from high to low), another where we remove cited sources at random. We then construct the reference overlap bibliographic coupling network and inspect its connectivity properties at regular intervals, as done in the previous subsection.

## 5) Results

We start by providing an overview of the (reference overlap) bibliographic coupling networks of the ten specialisms under consideration, in Table 4, considering data covering 5 years: 2011 to 2015 included, hence the partial drop in the number of articles (nodes).

| Specialism / Statistic | A1-ec_hist | A2-hist_sci | B1-NIPS | B2-ACL | C1-icarus | C2-JCAP | D1-classics | D2-eng_lit | E1-neuron | E2-MBE |
|---|---|---|---|---|---|---|---|---|---|---|
| # nodes | 812 | 1652 | 1843 | 1410 | 2088 | 2834 | 1313 | 918 | 1508 | 1324 |
| of which isolated | 5 | 40 | 7 | 2 | 4 | 4 | 151 | 58 | 6 | 2 |
| # edges | 16'555 | 52'518 | 82'792 | 110'230 | 134'428 | 396'347 | 23'320 | 5456 | 51'589 | 114'803 |
| density | 0.05 | 0.038 | 0.049 | 0.111 | 0.062 | 0.099 | 0.027 | 0.013 | 0.045 | 0.13 |
| diameter | 5 | 7 | 5 | 5 | 5 | 6 | 15 | 11 | 6 | 5 |
| **global clustering** | 0.37 | 0.38 | 0.36 | 0.44 | 0.39 | 0.51 | 0.32 | 0.26 | 0.29 | 0.44 |
| **modularity** | 0.38 | 0.35 | 0.39 | 0.33 | 0.45 | 0.31 | 0.32 | 0.5 | 0.37 | 0.24 |

*Table 4: General statistics for the reference overlap networks of the ten specialisms under consideration. Informally: an isolated node is one without edges; the density of the graph is the proportion of existing edges over the maximum possible given the amount of nodes; the diameter is the longest shortest path between any two nodes in the graph; the global clustering of the graph is the number of existing triangles of nodes (or closed triplets) over the maximum possible; the modularity, calculated using the fast greedy approach, gives an idea of how well the network can be partitioned into clusters (the higher the better). For formal definitions refer to Newman [2010].*

It is worth noting that, under some aspects, specialisms in history and literature stand out. Namely, having fewer edges, a higher number of isolated nodes with no edges, and an often lower density. This highlights from the very beginning that their networks are less well connected. Literature specialisms seem particularly weak in this respect, also considering their higher diameter, meaning that the distances in the network can be longer there. Under other aspects, such as modularity – or the quality of a network partition into clusters – there is less of a difference.

We compare next the reference and text similarity networks (built using Eq. 1 and 2 respectively). We group humanities specialisms (A and D) and science specialisms (B, C and E), following the hypothesis that the humanities specialisms are more rural, and science specialisms more urban. Consider first Equation 3. In Figure 2 we plot $c(t)$ on the y axis versus $t$ on the x axis, averaging results over the humanities and the sciences, for both networks. In Figure 3 we give the same results, averaged over every specialism instead. Consider next Equation 4. In Figure 4 we plot $g(t)$ on the y axis versus $t$ on the x axis, with the same set-up as in Figure 2. Results for the size of the giant component are coherent with those for connectivity.

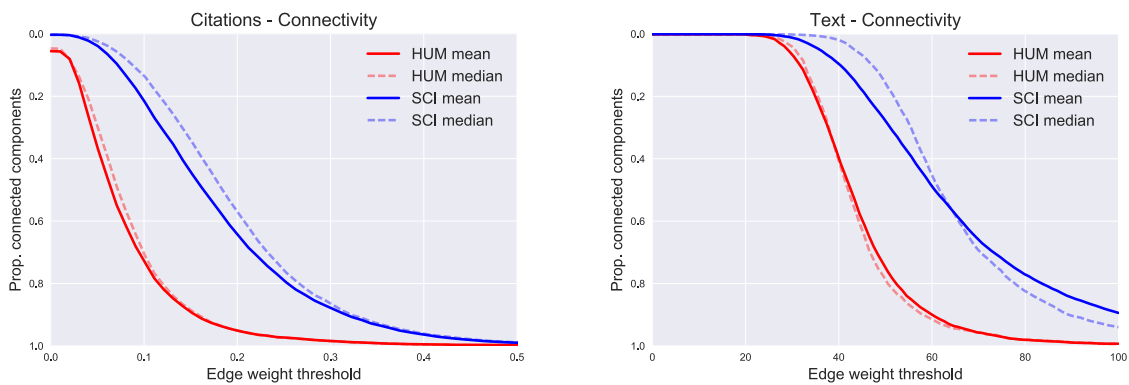

*Figure 2: Mean and median connectivity of the reference (left) and text similarity (right) networks, grouped into the humanities (red/grey) and the sciences (blue/dark grey).*

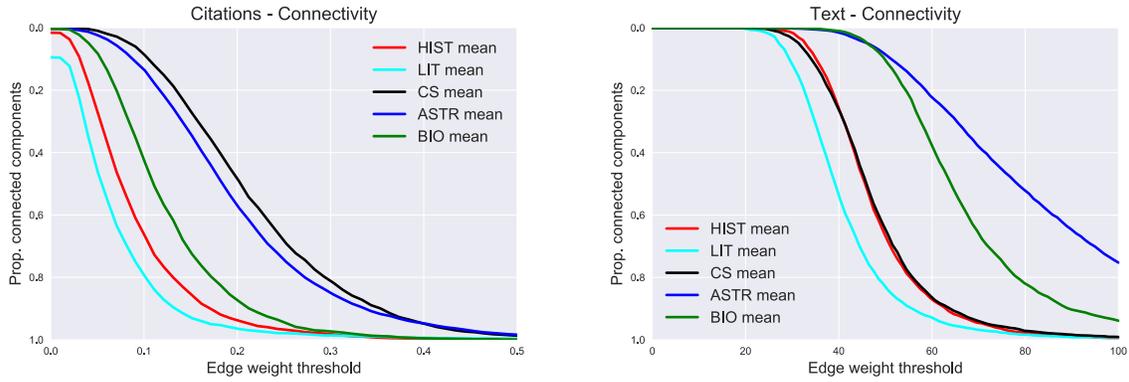

*Figure 3: Mean and median connectivity of the reference (left) and text similarity (right) networks, by specialism. Legend: HIST: A, LIT: D, CS: B, ASTR: D, BIO: E.*

Our results clearly highlight a lower overall connectivity for specialisms in the humanities, both over reference and textual similarities. Individually, specialisms behave differently. Astrophysics (C) has both high reference and textual similarities, while computer science (B) has higher reference and lower textual similarity (identical to history), biology (E) has higher textual and lower reference similarity, being closer to history than astrophysics in this respect. Nevertheless, all scientific specialisms have higher connectivity than specialisms in the humanities, across both similarity measures, with history presenting slightly higher similarity than literature. This result indicates that research topics are finer-grained in the humanities than in the sciences, as discussed in hypothesis 1, and this is consistent both with respect to reference overlap and textual similarity over titles and abstracts. However, the substantial observed variety across specialisms indicates that their cognitive and social structures are likely not fully explained within the rural and urban analogy.

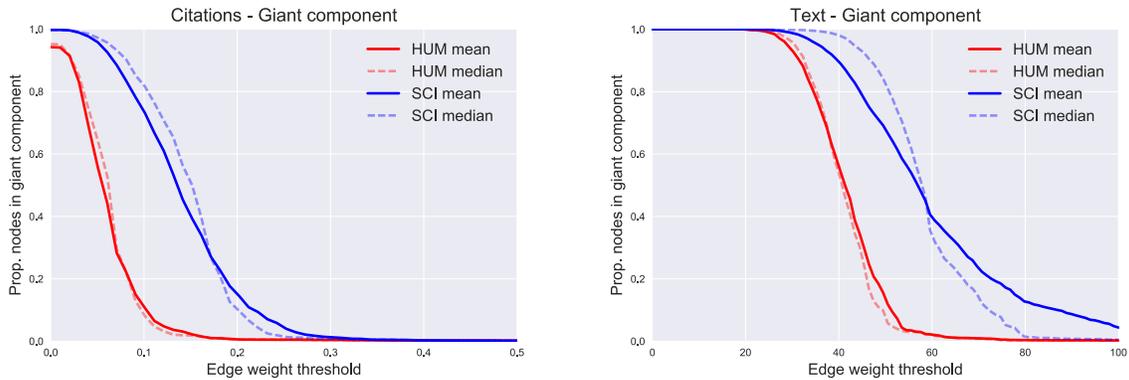

*Figure 4: Mean and median size of the giant component of the reference (left) and text similarity (right) networks, grouped into the humanities (red/grey) and the sciences (blue/dark grey).*

Given hypothesis 1 (i.e. urban specialisms maintain a higher connectivity, due to the presence of a lower number of larger topics), hypothesis 2 follows if urban specialisms also have more authors overall and more co-authorships, at the same weight threshold. Urban specialisms in our dataset indeed possess a higher number of unique authors (and a higher average number of co-authors per publication, cf. Table 1), thus hypothesis 2 follows immediately. This is because larger topics contain necessarily more authors than in rural specialisms, giving confirmation to the higher people-to-problem ratio of urban specialisms. In Table 5 we report the average and median size of the connected components with more than one node, and the corresponding people-to-problem ratio (number of unique authors per connected component), at different weight thresholds over the reference overlap network. Both the size of connected components and the people to problem ratio possess very skewed distributions. At the same time the size of topics and, especially, the people to problem ratio are sensibly higher for scientific specialisms, as expected.

| Specialism / Statistic | Threshold $t=0.1$ | | Threshold $t=0.2$ | | Threshold $t=0.3$ | |
| --- | --- | --- | --- | --- | --- | --- |
| | Topic size | p-t-p | Topic size | p-t-p | Topic size | p-t-p |
| A1 Economic History | 4 (2) | 5.9 (4) | 2.3 (2) | 3.5 (3) | 2 (2) | 2.9 (3) |
| A2 History of Science | 7.7 (2) | 7.9 (2) | 2.8 (2) | 2.9 (2) | 2.4 (2) | 2.2 (2) |
| B1 NIPS | 93.4 (2) | 159.5 (5) | 6.6 (2) | 14.9 (7) | 3 (2) | 6.9 (6) |
| B2 ACL | 100.8 (2) | 164.4 (6) | 10.3 (2) | 20.3 (6) | 3.8 (2) | 9.2 (6) |
| C1 Icarus | 69.1 (2) | 164.8 (10) | 5.4 (3) | 19 (11) | 2.8 (2) | 9.6 (7) |
| C2 JCAP | 34.2 (2) | 79.1 (5.5) | 6.2 (2) | 15.6 (6) | 3.1 (2) | 7 (5) |
| D1 Classics | 4.8 (2) | 4.4 (2) | 2.6 (2) | 2.3 (2) | 2.6 (2) | 2.4 (2) |
| D2 English Literature | 3.4 (2) | 3.6 (2) | 2.3 (2) | 3 (2) | 2.2 (2) | 3.2 (2) |
| E1 Neuron | 14.8 (2) | 89.3 (20) | 3 (2) | 19 (13) | 2.5 (2) | 15.6 (11) |
| E2 MBE | 8.6 (2) | 35.2 (12) | 2.8 (2) | 9.6 (7) | 2.4 (2) | 9.1 (5) |

*Table 5: Size of connected components mean (median) and people-to-problem ratio mean (median) for the reference overlap networks, calculated at different thresholds considering components with more than one node. By people with consider unique authors active in the component. An author can be active in more components via multiple publications.*

We further conducted the same experiments on all journals part of specialisms A and D individually (i.e. over networks of articles from the same journal only), to verify whether defining a specialism as an aggregation of articles from many journals would not artificially reduce the overall connectivity of the network. Indeed, no journal taken individually presents results markedly different from the overall trend of the respective specialism, thus we conclude that the lower overall connectivity in the humanities specialisms is not an artefact of journal aggregation. We omit these results for brevity.

The underlying process described by equations 3 and 4 is illustrated in Figure 5, for the case of specialisms B1 (NIPS) and D1 (Classics), considering increasing *t* (0.1, 0.2 and 0.3, left to right). It is possible to appreciate how the NIPS specialism is not only denser at low *t*, but is also maintaining larger connected components at higher thresholds, according to hypothesis 1 and as detailed in Table 4.

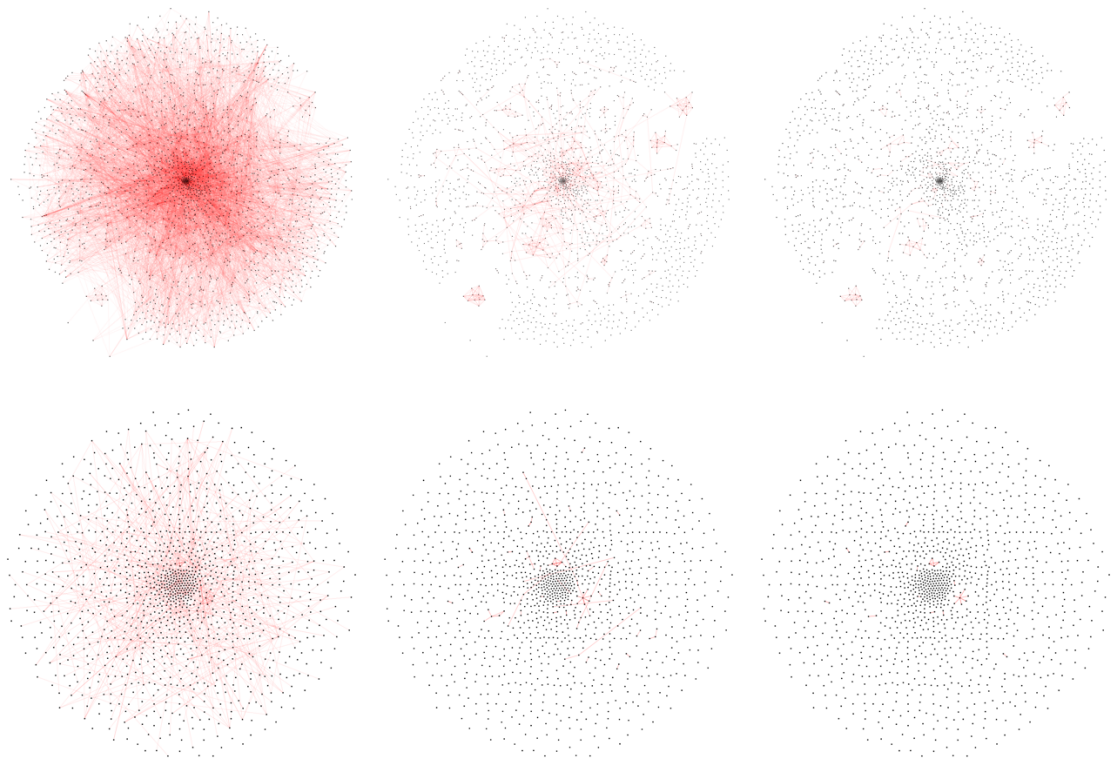

*Figure 5: Illustration of two reference overlap networks during the process of removal of edges below a given threshold. The same network layout is kept in all figures (using Force Atlas 2 in linlog mode from Gephi 0.9.1 [Bastian et al., 2009; Jacomy et al., 2014]). Above: B1 NIPS. Below: D1 Classics. Thresholds: 0.1 (left), 0.2 (center), 0.3 (right). The NIPS network presents several connected components even at relatively high thresholds, while the Classics network becomes almost disconnected already at threshold 0.2.*

Moving to consider the reliance of specialisms on their cited sources, we show results in Figure 6 for connectivity and Figure 7 for the giant component, averaging as before over the humanities and the sciences. In both cases, a process of removal in order of received citations is compared with one where edges where removed at random.

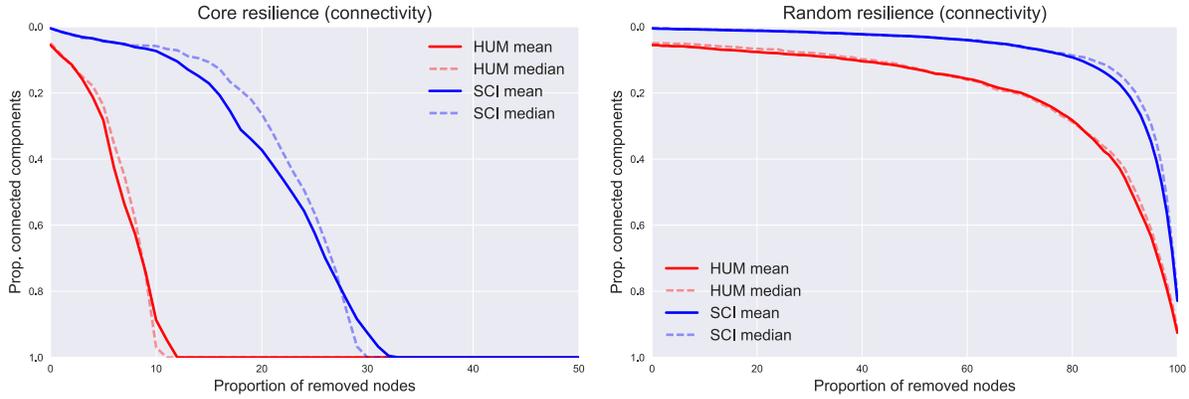

*Figure 6: Connectivity of the reference similarity networks to the removal of highly cited sources first (left) or at random (right), divided in the humanities (red/grey) and the sciences (blue/dark grey). The proportion of removed nodes is in %, thus 10 means 10%.*

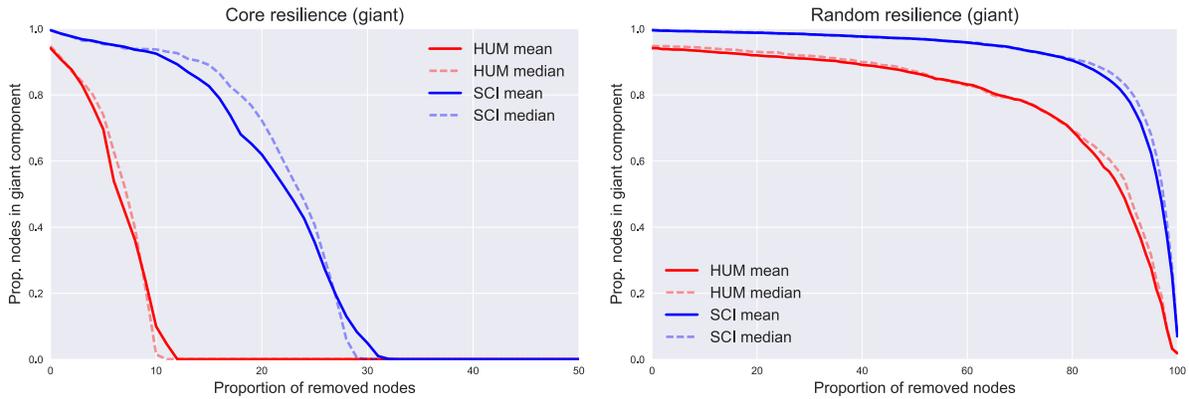

*Figure 7: Changing size of the giant component of the reference similarity networks to the removal of highly cited sources, divided in humanities (red/grey) and sciences (blue/dark grey). The proportion of removed nodes is in %, thus 10 means 10%.*

Two results emerge. Firstly, scientific specialisms are more reliant on core literature both at the specialism level and at the topic level, witness the higher resilience of their bibliographic coupling networks to the removal of the core literature at all stages. Secondly, the humanities present less well-connected bibliographic coupling networks in general, as shown by their lower connectivity apparent at all stages of the random removal process. To be sure, this phenomenon is much stronger in literature than history. We might conclude that the humanities possess a more fragmented intellectual base in terms of reference overlap, and in particular share fewer core sources at the specialism level, which in part contradicts what stated in hypothesis 6. The effect of the removal on the size of the giant component also suggests that the core literature is more substantially shared at the specialism level in the sciences, and not just within larger topics. These results are even stronger for literature, a pure humanities field, than history, which at times borders the social sciences—and especially so economic history. If rural specialisms are fragmented into small topics, as shown, and at the same time possess fewer core sources, it follows that their fragmentation is in part due to more focused, topic-specific or unique, non-overlapping referencing behaviour. But, also, the within topics connectivity is higher in science specialisms.

## 6) Discussion

The aim of our analysis was to offer a possible quantitative operationalization of Becher & Towler's [2001] conceptualization of the social and cognitive structure of research specialisms as rural or urban. According to this conceptualization rural specialisms show, in comparison to urban ones, (1) a higher number of smaller topics being researched, (2) a lower people-to-problem ratio, (3) longer publications, that (4) contain more references, and also show (6) to share comparatively more references across topics than within, and (8) to have less co-authorships.

We have offered an operationalization of this conceptualization of the social and cognitive structure of research specialisms by comparing the textual and reference connectivity among publications within ten humanities and science specialisms. We use publication venues (journals) to proxy specialisms and well-connected clusters in the bibliographic coupling network of publications to proxy topics. Considering reference connectivity first, and focusing on hypothesis 1 and 6, we find that hypothesis 1 is confirmed at the specialism level, as science specialisms are overall better connected than in the humanities, with some disciplinary variations. Similarly using textual connectivity, we see a stronger connectivity in the sciences, especially at smaller topics, likely the effect of technical jargon and a higher degree of specialization. An exception at the textual level is computer science, behaving on par with history. With respect to hypothesis 2 we find strong supporting evidence for a considerably higher people-to-problem ratio in science specialisms, or the number of active authors per topic. However, topics are not that easily defined, and we do not find many distinct clusters in any specialism. This leads us to argue that within the sciences, specialisms are comparatively well-connected at both the level of general, larger topics and the level of smaller, tighter ones, but that the distinction between these two levels of the cognitive structure is not as clear as Becher and Towler suggest. Within the humanities we find a comparatively lower connectivity also on the level of the specialism. This means that we do not find any evidence for the idea that humanities scholars tend to cite more broadly to establish an intellectual base for their contribution within the specialism as a whole.

In light of these findings we suggest to re-assess the use of the rural versus urban conceptualization. Despite the fact that some already established elements of this conceptualization find confirmation in our analysis (e.g. length of publications, number of active authors, people-to-problem ratio), the specific cognitive structure of specialisms and topics is only partially reproduced. Rather, we find that science specialisms show an overall cohesion that suggests that scholars work in a particular paradigm in which topics are not necessarily clearly distinguished. The overall fragmentation of the humanities specialism suggests instead a less unified cognitive structure, at least to the extent to which this is articulated through reference lists and textual similarity. It is very well possible that in these humanities fields a particular paradigm is dominant without scholars having to articulate it or having to make reference the historical sources that lay at the basis of this paradigm. To further uncover these links, it might be necessary to consider different data sources for which inspiration might be found in previous work on invisible colleges [Cronin, 1982]. It might also be that the limited size of the corpus (both in number of articles and over time) limits our ability to find evidence for particular topics that might exist across a broader spectrum of publications. Finally, the journal article might assume different roles in the humanities, for example as a more specialized form of publication, as monographs play a more important role there than in the sciences [Thompson, 2002; Williams et al., 2009]. Indeed, our results are limited by the use of journal articles as publication data and might change should we consider other publication typologies such as books. Future research might also explore what type of sources are widely shared within specialisms in the sciences, for instance whether these are primarily theoretical or methodological, and whether humanities scholars signal their theoretical or methodological position in a way that cannot be traced bibliometrically.

An important question this study raises is related to the evaluation of research in the humanities. The differences we observed in the study with respect to publication cultures between science and humanities specialisms have been illustrated before [van Leeuwen et al., 2016; Kulczycki et al., 2018] by showing the diversity of output types coming from the humanities versus the more monotone rhythm of journal publishing from natural sciences. Previous literature has also considered citation patterns over time, connected to the speed of knowledge becoming obsolete [Colavizza, 2017], and the way references are used [Knievel & Kellsey, 2005]. Our results indicate that reference behaviour in the humanities is very different compared to the sciences, both at the level of a specialism and a topic. Without a more in-depth sociological understanding of these differences it is not advisable to use citation measures in an evaluative fashion to compare humanities and sciences specialisms.

Crucial in the future of the (quantitative) study of the social and cognitive structure of (humanities) fields, as well as their evaluation, is a far stronger integration of empirical operationalizations and theory-development. To take just one example, our results indicate that Whitley's [1984] conceptualization of the cognitive and social structure of fields might shed light on the structures found in our empirical study. This work has, despite its influence and similar to Becher and Trowler, had little empirical validation. Drawing on a variety of data sources, methods and analytical angles will be necessary to establish the variety of social and cognitive structures of fields as well as their historical development. A task for which scientometrics, the sociology and history of science as well as science and technology studies will need to be combined; a task which is, according to us, long overdue.

# Bibliography

Bastian M., Heymann S. and Jacomy M. (2009). Gephi: an open source so ware for exploring and manipulating networks. In *International AAAI Conference on Weblogs and Social Media*. San Jose, CA.

Becher T. (1989). *Academic tribes and territories: intellectual enquiry and the culture of disciplines*. Open University Press, Milton Keynes.

Becher T. and Trowler P. (2001, 2nd ed.). *Academic tribes and territories: intellectual enquiry and the culture of disciplines*. Open University Press, Philadelphia.

Boyack K. W., Newman D., Duhon R. J., Klavans R., Patek M., Biberstine J. R., Schijvenaars B., Skupin A., Ma N. and Börner K. (2011). Clustering more than two million biomedical publications: comparing the accuracies of nine text-based similarity approaches. *PLoS ONE*, 6(3): e18029.

Boyack K. W., Klavans R. and Börner K. (2005). Mapping the backbone of science. *Scientometrics*, 64(3): 351–374.

Börner K. (2010). *Atlas of science: Visualizing what we know*. MIT Press, Cambridge MA.

Börner K., Chen C. and Boyack K. W. (2003). Visualizing knowledge domains. *Annual Review of Information Science and Technology*, 37(1): 179–255.

Chen C. (2017). Science Mapping: A Systematic Review of the Literature. *Journal of Data and Information Science,* 2(2): 1–40.

Colavizza G. (2017). The core literature of the historians of Venice. *Frontiers in Digital Humanities*, 4(14).

Colavizza G., Boyack K. W., van Eck N. J. and Waltman L. (2018). The closer the better: similarity of publication pairs at different co-citation levels. *Journal of the Association for Information Science and Technology*, 69(4): 600–609.

Cole S. (1983). The Hierarchy of the Sciences? *American Journal of Sociology*, 89(1): 111–139.

Cronin B. (1982). Invisible colleges and information transfer. A review and commentary with particular reference to the socia sciences. *Journal of Documentation*, 38(3): 212–236.

Fanelli D. and Glänzel W. (2013). Bibliometric Evidence for a Hierarchy of the Sciences. *PLoS ONE,* 8(6): e66938.

Fuchs S. (1993). A sociological theory of scientific change. *Social Forces*, 71(4): 933–953.

Gläser J. and Laudel G. (2016). Governing science. *European Journal of Sociology*, 57(01): 117–168.

Hammarfelt B. (2011). Interdisciplinarity and the intellectual base of literature studies: citation analysis of highly cited monographs. *Scientometrics*, 86(3): 705–725.

Hammarfelt B. (2012). Harvesting footnotes in a rural field: citation patterns in Swedish literary studies. *Journal of Documentation*, 68(4): 536–558.

Hammarfelt B. (2016). Beyond coverage: Toward a bibliometrics for the humanities. In: M. Ochsner, S. E. Hug and H.-D. Daniel (eds.), *Research Assessment in the Humanities. Towards Criteria and Procedures,* p. 115–131. Springer International Publishing, Cham.

Harzing A.-W. and Alakangas S. (2016). Google Scholar, Scopus and the Web of Science: A Longitudinal and Cross-Disciplinary Comparison. *Scientometrics,* 106(2): 787–804.

Jacomy M., Venturini T., Heymann S. and Bastian M. (2014). ForceAtlas2, a continuous graph layout algorithm for handy network visualization designed for the Gephi software. *Plos One,* 9: e98679.

Kessler M. M. (1963). Bibliographic coupling between scientific papers. *American Documentation,* 14: 10–25.

Knievel J. E. and Kellsey C. (2005). Citation Analysis for Collection Development: A Comparative Study of Eight Humanities Fields. *The Library Quarterly,* 75(2): 142–68.

Kulczycki E., Engels T. C., Pölönen J., Bruun K., Dušková M., Guns R., Nowotniak R., Petr M., Sivertsen G., Starčič A. I. and Zuccala, A. (2018). Publication patterns in the social sciences and humanities: evidence from eight European countries. *Scientometrics*, 116(1): 1–24.

Larivière V., Archambault É. and Gingras Y. (2008). Long-term variations in the aging of scientific literature: From exponential growth to steady-state science (1900–2004). *Journal of the Association for Information Science and Technology*, 59(2): 288–296.

Leydesdorff L. (1989). The relations between qualitative theory and scientometric methods in science and technology studies: Introduction to the topical issue. *Scientometrics*, 15(5-6): 333–347.

Leydesdorff L. and Rafols I. (2009). A global map of science based on the ISI subject categories. *Journal of the Association for Information Science and Technology*, 60(2): 348–362.

Leydesdorff L., Hammarfelt B. and Salah A. (2011). The structure of the Arts & Humanities Citation Index: A mapping on the basis of aggregated citations among 1,157 journals. *Journal of the Association for Information Science and Technology*, 62(12): 2414–2426.

Luukkonen T. (1997). Why has Latour's theory of citations been ignored by the bibliometric community? Discussion of sociological interpretations of citation analysis. *Scientometrics,* 38(1): 27–37.

Merton R. (1974). *The sociology of science: theoretical and empirical investigations*. University of Chicago Press, Chicago.

Morris S. and Van der Veer Martens B. 2008. Mapping Research Specialties. *Annual Review of Information Science and Technology,* 42(1): 213–295.

Newman M. (2010). *Networks: an introduction*. Oxford University Press, Oxford.

Price D. De Solla (1970). Citation measures of hard science, soft science, technology, and nanoscience. In C. E. Nelson and D. K. Pollock (eds.), *Communication among scientists and engineers*, p. 3–22. Heath Lexington Books, Lexington Mass.

Storer N. (1967). The hard sciences and the soft: some sociological observations. *Bulletin of the Medical Library Association*, 55(1): 75–84.

Spark Jones K., Walker S. and Robertson S. E. (2000a). A probabilistic model of information retrieval: development and comparative experiments. Part 1. *Information Processing and Management,* 36(6): 779–808.

Spark Jones K., Walker S. and Robertson S. E. (2000b). A probabilistic model of information retrieval: development and comparative experiments. Part 2. *Information Processing and Management,* 36(6): 809–840.

Thompson J. W. (2002). The Death of the Scholarly Monograph in the Humanities? Citation Patterns in Literary Scholarship. *Libri,* 52(3): 121–136.

Tonnies F. (1957). *Community and Society.* Harper Torchbook, New York.

Trowler P., Saunders M. and Bamber V. (eds.) (2012). *Tribes and territories in the 21st century: Rethinking the significance of disciplines in higher education.* Routledge, London.

Trowler P. (2014). Depicting and researching disciplines: Strong and moderate essentialist approaches. *Studies in Higher Education*, 39(10): 1720–1731.

Tsai C., Corley E. and Bozeman B. (2016). Collaboration experiences across scientific disciplines and cohorts. *Scientometrics*, 108(2): 505–529.

van Leeuwen T. N., van Wijk E. and Wouters P. F. (2016). Bibliometric analysis of output and impact based on CRIS data: A case study on the registered output of a Dutch university, *Scientometrics*, 106(1): 1-16.

Whitley R. (1984). *The intellectual and social organization of the sciences*. Oxford University Press, Oxford.

Williams P., Stevenson I., Nicholas D., Watkinson A. and Rowlands I. (2009). The Role and Future of the Monograph in Arts and Humanities Research. *Aslib Proceedings,* 61(1): 67–82.

Wouters P. and Leydesdorff L. (1994). Has Price's dream come true: Is scientometrics a hard science? *Scientometrics*, 31(2): 193–222.

Wyatt S., Milojevic S., Park H. and Leydesdorff L. (2016). The intellectual and practical contributions of scientometrics to STS. In U. Felt, R. Fouché, C. A. Miller and L. Smith-Doerr (eds.), *The handbook of science and technology studies*, p. 87–112. MIT Press, Cambridge MA.

Zuckerman H. and Merton R. (1973). Age aging and age structure in Science. In: R. Merton and N. Storer *The Sociology of Science. Theoretical and Empirical Investigations*, p. 497–539. University of Chicago Press, Chicago.